\documentclass[a4paper]{jpconf}
\usepackage{graphicx}

\newcommand{\SgrA}{Sgr~A$^\ast$}
\newcommand{\MBH}{\ensuremath{M_{\bullet}}}
\newcommand{\Msun}{\ensuremath{{\rm M}_{\odot}}}

\newcommand{\Nstar}{\ensuremath{N_{\ast}}}

\newcommand{\pc}{\ensuremath{\mathrm{pc}}}

\newcommand{\peryr}{\ensuremath{\mathrm{yr}^{-1}}}

\newcommand{\trlx}{\ensuremath{t_{\rm rlx}}}

\newcommand{\Rperi}{\ensuremath{R_{\rm peri}}}

\newcommand{\NBFOUR}{\sc Nbody4}

\newcommand{\tento}[2]{\ensuremath{{#1}\times 10^{#2}}}

\begin{document}

\title{Models of mass segregation at the Galactic centre}

\author{Marc Freitag$^{1,2}$, Pau Amaro-Seoane$^3$  and Vassiliki Kalogera$^1$}

\address{$^1$Department of Physics and Astronomy, 
Northwestern University, Evanston, IL 60208, USA}
\address{$^2$Institute of Astronomy, University of Cambridge, Madingley Road, CB3~0HA Cambridge, UK}
\address{$^3$Max Planck Institut f\"ur Gravitationsphysik, D-14476 Potsdam, Germany}

\ead{freitag@ast.cam.ac.uk}

\begin{abstract}

We study the process of mass segregation through 2-body relaxation in
galactic nuclei with a central massive black hole (MBH). This study
has bearing on a variety of astrophysical questions, from the
distribution of X-ray binaries at the Galactic centre, to tidal
disruptions of main-sequence and giant stars, to inspirals of compact
objects into the MBH, an important category of events for the future
space borne gravitational wave interferometer LISA. In relatively
small galactic nuclei, typical hosts of MBHs with masses in the range
$10^4-10^7\,\Msun$, the relaxation induces the formation of a steep
density cusp around the MBH and strong mass segregation. Using a
spherical stellar dynamical Monte-Carlo code, we simulate the
long-term relaxational evolution of galactic nucleus models with a
spectrum of stellar masses. Our focus is the concentration of stellar
black holes to the immediate vicinity of the MBH. Special attention is
given to models developed to match the conditions in the Milky Way
nucleus.
\end{abstract}

\section{Method, physics and initial conditions}

This contribution summarises our recent series of simulations of
galactic nuclei aimed at the investigation of mass segregation
\cite{FASK06}.

We used a stellar dynamics code based on the Monte Carlo (MC) scheme
pioneered by H\'enon
\cite{Henon71a,Henon71b}. This method offers a unique compromise
between physical realism and computational efficiency. Like in direct
$N-$body codes, the stellar system is represented as a set of
particles; this allows the implementation of a rich physics:
self-gravity, general stellar mass spectrum and velocity distribution,
2-body relaxation, collisions between stars, stellar evolution,
binaries (not included in the present code), tidal disruptions and
captures by the central MBH
(\cite{FB01a,FB02b,Freitag01,Freitag03,FreitagRB06,FreitagGR06} for a
description of the code and previous applications). Unlike $N-$body,
the MC method assumes spherical symmetry and dynamical
equilibrium. This makes it much faster; while the CPU time per
relaxation time increases like $N^3$ for $N-$body simulations, the MC
code has a $N\ln N$ scaling.

We performed about 90 different simulations, to
investigate the effects of various physical ingredients, assumptions
about their treatment, of the initial nucleus structure and to perform
some limited parameter-space exploration. For most models,
$\tento{4}{6}$ particles where used, requiring a few days of computing
time on a single-CPU PC. A few cases were computed with $10^6$ or
$\tento{8}{6}$ particles to establish that our results are not
strongly affected by the limitations in numerical resolution.

All runs included the effects of the gravity of a central MBH, of the
self gravity of the stars, of 2-body relaxation, treated in the
Chandrasekhar (diffusive) approximation, and of the tidal disruption
of main-sequence (MS) stars at the Roche limit around the MBH as well as direct
coalescence with the MBH for stars too compact to be tidally
disrupted. In most cases, stellar evolution was not included
explicitly; instead the stellar population consists, from the
beginning of the simulation, of a mixture of MS and compact remnants
corresponding to a single star formation episode that took place
10\,Gyr ago. In a few models, explicit stellar evolution was included
with all stars starting on the MS and turning into compact remnants
at the end of their MS life time. For simplicity, giants were
not considered because, as far as mass segregation is concerned, only
the mass of the star matters and the evolution of the stellar
distribution, being a relaxational process requires timescales much
longer than the duration of the giant phase. Stellar collisions and
large-angle gravitational deflections (not accounted for in the
diffusive treatment of relaxation) were considered in a small number
of models. We made no attempt to determine whether a given star-MBH
coalescence would occur as a gradual ``extreme-mass ratio inspiral''
(EMRI) detectable by LISA or a direct plunge through the horizon of
the MBH (see \cite{Sigurdsson03,Glampedakis05} for background
information and references on EMRIs and \cite{HB95,HA05} for the
importance of the distinction between EMRIs and plunges).

In all our runs the galactic nucleus is started as an $\eta$-model
\cite{Dehnen93,Tremaine94}, with a central power-law density
cusp, $\rho \propto R^{\eta-3}$ and steeper ``cut-off'' at large
radii, $\rho \propto R^{-4}$. In most cases, we used parameters (mass
of the MBH, stellar density around it, etc)  corresponding to the
stellar cluster around {\SgrA} at the centre of our Galaxy
\cite{SchoedelEtAl03,GhezEtAl05}. We did not try to
reproduce the very peculiar spatial and age distribution of the bright
IR stars observed within 1\,pc of {\SgrA}. In this work, we adopt the
position that these stars, useful as they are as probes of the
gravitational potential, are not representative of the overall stellar
population at the Galactic centre, assumed to be much older and
therefore amenable to our treatment. This defines a well-posed problem
which constitutes an interesting limiting case. Clearly other situations
have to be considered in future studies.

In addition to the {\SgrA} models, we followed the evolution of
galactic nuclei hosting an MBH with a mass $\MBH$ in the range
$10^4-10^7\,\Msun$. Based on a somewhat naive application of the
$M-\sigma$ relation \cite{TremaineEtAl02}, we scaled the size of the
stellar cluster according to $R_{\rm nucl}\propto \MBH^{0.5}$ where
$R_{\rm nucl}$ is any characteristic length of the stellar
distribution. We have considered two families of
models; one with $\eta=2.0$, the other with
$\eta=1.5$. The interval in $\MBH$ was chosen mostly to cover the
values that should yield gravitational wave signals in LISA band when
a compact star inspiral into the MBH. The present study is a first
step towards more robust determinations of the rate and characteristic
of such EMRIs. This range of models also covers systems that are both
large enough (in terms of the number of stars) to be amenable to
treatment with the MC method and small enough for relaxational effects
to play a significant role over some 10\,Gyr.

To ensure that the MC code, based as it is on a number of simplifying
assumptions, yield correct results, we carried out a number of
comparisons with simulations performed with the highly accurate (but much
more computationally demanding) direct-summation {\NBFOUR} code
\cite{Aarseth99,Aarseth03b}. In particular, we compared with results obtained by
Baumgardt et al.\ \cite{BME04a} and performed a new {\NBFOUR} simulation of a
two-component model with a central massive object using 64\,000
particles. On the GRAPE hardware at disposal not more than $\sim 10^5$
particles can be used; hence it is not yet possible to simulate a
system with a realistic mass function using direct $N-$body but this
2-component toy model demonstrated for the first time in a direct
fashion that the MC code treats mass segregation around a MBH very
satisfactorily.

\section{Results}

\begin{figure}[h]
\includegraphics[bb=27 155 583 698,clip,width=20pc]{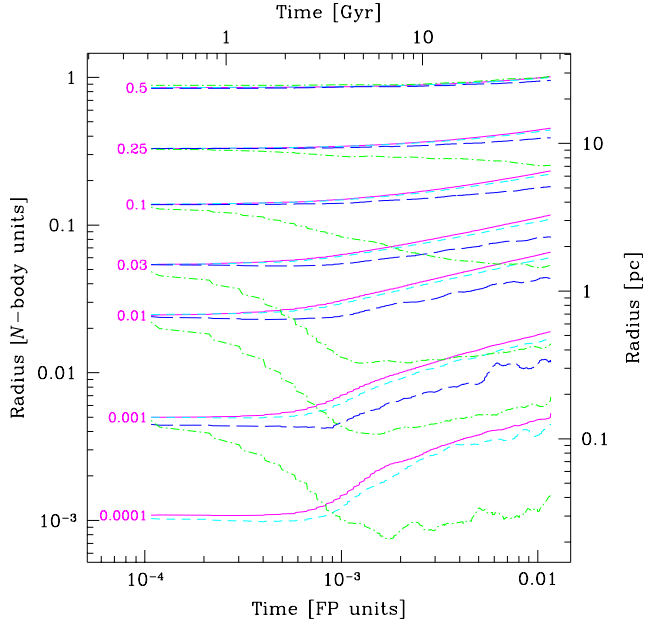}
\hspace{2pc}
\begin{minipage}[b]{16pc}
\caption{\label{fig:LagStd} Evolution of Lagrange radii for a ``standard'' {\SgrA} nucleus
  model.  We plot the evolution of the radii of spheres that enclose
  the indicated fractions of the mass of various stellar
  species. Solid lines are for MS stars, short-dashed lines
  for white dwarfs, long-dashed lines for neutron stars and
  dash-dotted lines for stellar BHs. For this model, a 10\,Gyr-old,
  non-evolving stellar population was used and the central MBH has an
  initial mass of $\MBH=\tento{3.6}{6}\,\Msun$. It grows only very little
  through tidal disruptions and stellar coalescences.}
\end{minipage}
\end{figure}

\begin{figure}[h]
\includegraphics[bb=27 155 583 698,clip,width=20pc]{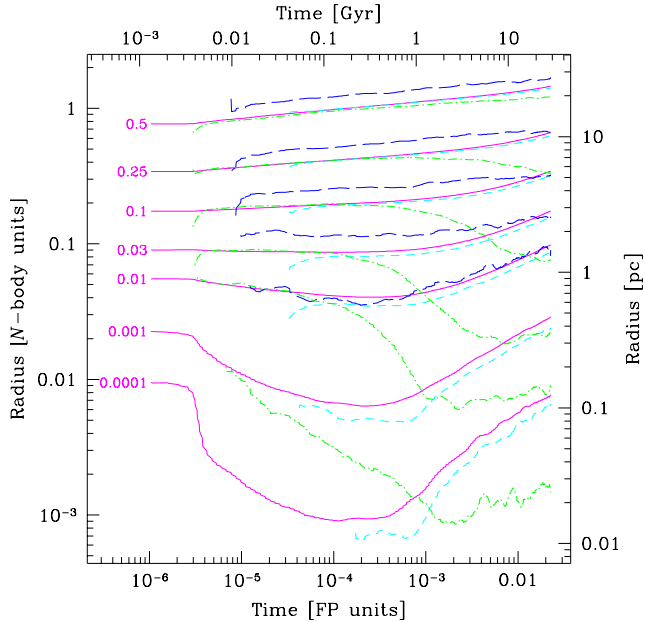}
\hspace{2pc}
\begin{minipage}[b]{16pc}
\caption{\label{fig:LagSE} Same as Fig.~\ref{fig:LagStd} but for a {\SgrA} model with 
all stars on the MS and $\MBH=1000\,\Msun$ at start. The
MBH grows by accreting an {\em ad hoc\/} fraction of the gas emitted
by stellar evolution to reach $\MBH\simeq\tento{3.9}{6}\,\Msun$ after
10\,Gyr. The structure of the nucleus after $\sim 5\,$Gyr is similar
to that of the case without stellar evolution. The initial size of the
cluster was adjusted to obtain a reasonably good fit to the observed
stellar mass profile around {\SgrA} at $t=10\,$Gyr.}
\end{minipage} 
\end{figure}

For the {\SgrA} models, our main results are the following. In all
cases, the stellar BHs, being the most massive objects (with a fixed
mass of $10\,\Msun$ or a range of masses, depending on the model),
segregate to the central regions. This segregation takes about 5\,Gyr
to complete. The nucleus then enters a second evolutionary phase which
is characterised by the overall expansion of the central regions,
powered by the accretion of stellar mass (of very negative energy)
on to the MBH. Although all species participate in the expansion, mass
segregation continues in a relative fashion, as the system of BHs
expands slower than the other components. The structure of the nucleus
at distances from the centre larger than $\sim 10\,$pc is unaffected
by relaxation over a Hubble time. The evolution of two {\SgrA} models
is shown in Figures~\ref{fig:LagStd} and \ref{fig:LagSE}.

BHs dominate the mass density within $\sim 0.2\,\pc$ of the MBH but we
do not find them to be more numerous than MS stars in any region we
can resolve (down to a few mpc, at $t=10\,$Gyr). Estimating the
exponent for the density cusp the BHs form, $\rho
\propto R^{-\gamma}$, is difficult because of numerical noise but, 
in most cases, $\gamma$ is compatible with the Bahcall-Wolf value
$\gamma=1.75$ \cite{BW76,BW77}. In contrast, the less massive objects,
such as MS stars, form a cusp with $\gamma$ generally in the range
$1.3-1.4$ which is significantly lower than the value of $1.5$
predicted by
\cite{BW77}. This is also found in the 2-component $N-$body
simulation. After 5-10\,Gyr of evolution, we find of order
$\tento{2-3}{3}$, $\tento{6-8}{3}$ and $\tento{2}{4}$ stellar BHs
within 0.1, 0.3 and 1\,pc of the centre, respectively. About $10^4$
BHs coalesce with the MBH during a Hubble time. Using the formalism of
the dynamical friction for objects on circular orbits in a fixed
stellar background is an easy alternative for estimating the
concentration of massive objects in the central regions. However, for
stellar BHs, although this approach offers a qualitatively correct
picture, it overpredicts the effectiveness of mass segregation. In the
case of a model with $\eta=1.5$ (whose relaxation time does not
increase towards the centre), this yields too large a number of BHs
accreted by the MBH and {\it too few} (by a factor 3--5) being present
within the inner 1\,pc after some 10\,Gyr.

All types of objects lighter than the BHs, including the neutron stars, are
pushed away from the central regions. Using the observed distribution
of these objects to infer the presence of segregated BHs \cite{CG02}
does not seem to be possible, though because, in the absence of BHs,
it would take the neutron stars more than 10\,Gyr to form a Bahcall-Wolf Cusp of
their own, even without natal kick.

These results are not significantly affected by stellar collisions,
large-angle scatterings or the initial $\eta$ value. We also
considered three different prescriptions for the masses (and types) of
compact remnants and found no strong variations in the simulation
outcomes. Most interestingly, an alternative model in which stellar
evolution was included and the central MBH was grown from an IMBH seed
(by accretion of an {\it ad hoc} fraction of the mass lost by stars
when they turn into compact objects) yields basically the same
structure of mass segregation (and same rates of coalescences and
tidal disruptions) at $t\simeq 10\,$Gyr (see
Fig.~\ref{fig:LagSE}). These findings suggest that our main results
are not very sensitive to the special ``initial conditions'' used, as
long as they are fine-tuned to produce at $t=10\,Gyr$ a given MBH mass
and stellar mass within $\sim 1\,\pc$ of the MBH. However, it would be
instructive to consider a larger variety of models in future work,
including some with extended period of stellar formation. Our present
assumption of a single burst of stellar formation maximises the number
fraction of stellar BHs and the time available for mass segregation.

We find rates of tidal disruptions and coalescences with the MBH of
$\sim \tento{4}{-5}\peryr$ and $\sim \tento{3}{-5}\peryr$,
respectively, at $t=10\,$Gyr. When stellar collisions are allowed,
they occur at a rate about 10 times lower. On average, each MS-MS
collision releases about $0.01\,\Msun$ of gas; if collisions between
MS stars and compact remnants lead to complete
disruption of the MS star, collisions (of all types) yield an average
of $0.05-0.06\,\Msun$ per event.

When large-angle scatterings are explicitly included (essentially as a
special case of collisions), they are found to have only little impact
on the rate of tidal disruptions or coalescences with the MBH. A
stellar BH is about 10 times more likely to be swallowed by the MBH
than to be ejected from the nucleus. In contrast with this, in their
multi-mass $N-$body simulations, Baumgardt et al.\ \cite{BME04b} find
that all stellar BHs except one are ejected from the cluster and
ascribe this result to strong interactions with objects (generally
another stellar BH) deeply bound to the IMBH. These interactions are
likely to be ``resonant'', i.e., the three objects (including the
IMBH) form a strongly interacting, chaotic configuration for many
orbital times until one of the lighter objects is ejected (Baumgardt,
personal communication; see e.g.,
\cite{Hut93}). In principle, this mechanism can be included into 
the MC code by extending the loss-cone treatment used for tidal
disruptions and coalescences to interactions with the binary
consisting of the MBH and the most bound stellar object and resorting
to explicit integration of 3-body motion for close interactions
between the binary and a third object. However a
simple analysis based on extrapolation of the cross sections for
single-binary encounters \cite{HHMcM96} suggests that, in galactic
nuclei, such ejections are considerably less likely to occur than a
direct plunges through the MBH's horizon.

To first order, the evolution of galactic nuclei of different masses
(but same initial $\mu=\MBH/M_{\rm stars}$ and $\eta$ values, where
$M_{\rm stars}$ is the total stellar mass), can be obtained by
rescaling the mass, the size and the time units if stellar
evolution is negligible. The evolution timescale is set by the
relaxation time which increases approximately like $\trlx \propto
\MBH^{5/4}$ for our $R_{\rm nucl}\propto
\MBH^{0.5}$ assumption. This means that small nuclei are affected
by faster segregation and MBH-driven expansion and 
may have been significantly more compact in the past than observed
in the local universe. On the other hand, only very little
relaxational mass-segregation can have occurred in nuclei with MBHs
more massive than $\sim 10^7\,\Msun$ over a Hubble time. The evolution of the region very
close to the MBH cannot be scaled from one nucleus to another with
different mass and size, however because the processes of tidal
disruptions and coalescences with the MBH introduce physical
scales. In particular, we find that stellar BHs in more massive nuclei
experience less segregation than in smaller nuclei (at a given value
of $t/\trlx$). Also the rate of coalescences (in units of
$\Nstar/\trlx$, where $\Nstar$ is the total number of stars) is higher
in more massive nuclei. This is probably an effect of a larger
critical radius\footnote{The critical radius is essentially the
typical value of the semimajor axis for stars swallowed by the MBH.},
requiring a larger accretion rate to yield the same amount of energy
to power the expansion of the nucleus.

\section{Observational consequences and future developments}

Although we have not attempted a realistic modelling of the Galactic
centre, it is tempting to apply our results to one specific
observation of the {\SgrA} region. Using the {\em Chandra} X-ray
space-born observatory, Muno et al.\ \cite{MunoEtAl05} have detected 7
transient sources which appear to be much more concentrated around
{\SgrA} than the overall stellar population. Here we examine whether
this may be a direct consequence of mass segregation, if these sources
are all stellar BHs accreting from a lower-mass companion. We make the
strong assumption that these binaries are not formed or affected by
interactions with other stars such as 3-body binary formation, partner
exchange, ionisation, etc. Instead we consider that they just react to
2-body relaxation as point objects with a total mass approximated by
the mass of the stellar BH. If we pick up at random 7 sources with
projected distance from the centre smaller or equal to 23\,pc
according to the profile for stellar BHs found in our simulations, we
find that their distribution is at least as concentrated as the
observed one in 15\,\% of the cases. It is therefore at this point not
possible to exclude that the transients owe their peaked profile
purely to mass segregation but this seems somewhat unlikely.

As pointed out by Muno et al., the rate of binary
interactions should also increase steeply towards the centre and this
probably combines with mass segregation to produce the observed
distribution. The problem of binary
dynamics in the vicinity of a MBH is complicated by the fact that
there is no clear-cut definition of the hard-soft transition \cite{HH03}. The
Keplerian velocity dispersion increases virtually without bound when
one approaches the centre. This may affect a binary on an orbit of
relatively large semimajor axis $a$ around the MBH because 2-body
relaxation will cause the orbit to reach down to a value $\Rperi =
(1-e)a \ll a$ over a timescale of order
$\trlx\ln(1/(1-e))$ \cite{FR76}.

The most extreme type of dynamical interaction a binary can experience
is the tidal separation of its members if its orbit brings it within
$\sim a_{\rm bin}(\MBH/m_{\rm bin})^{1/3}$ of the MBH. Here $a_{\rm
bin}$ is the semimajor axis of the binary itself and $m_{\rm bin}$ its
mass. This process is of great interest by itself both as a way to
create ``hypervelocity stars'' and to deposit a star on a tight orbit
around the MBH \cite{Hills88,YT03,GPZS05,MFHL05,Pfahl05}.

To our knowledge, the complex question of binary dynamics in a
galactic centre has not been investigated yet. This is an ideal
subject for self-consistent stellar dynamical simulations of the sort
presented here but including binary processes. MC codes are
particularly well suited for following the evolution of large systems
with a significant fraction of binaries whose interaction can be
computed accurately by direct $3-$ and $4-$body integrations
\cite{GS03,FGR06,GFR06}.

The stellar BHs at the Galactic centre may have detectable
consequences even if they are single. If the motions of the S-stars
can be tracked with high enough a precision, an extended distribution
of non-luminous matter around the Galactic MBH should signal itself
through its effect on their orbits. Present-day observations are
insufficient to detect the slight Newtonian retrograde precession
induced by an extended ``dark cusp'' \cite{MouawadEtAl05}, but future
$30-100$\,m telescopes may allow to realise such a measurement and to
witness $\sim 3$ trajectory deflections caused by gravitational
encounters per year between any of $\sim 100$ monitored S-stars and a
stellar BH \cite{WMG05}. The compact stars can also collide and merge
with MS stars and giants, hence creating unusual objects, once
suggested to be the S-stars themselves \cite{Morris93}, or increase
the rotation rate of extended stars through multiple tidal
interactions \cite{AK01}. In this context, we note that our models
predict about $\tento{2-3}{4}$ collisions between MS
stars and stellar BHs and a similar number of MS--white-dwarf events
to occur over a Hubble time in a {\SgrA}-type nucleus. Collisions with
neutron stars are at least 10 times rarer. Finally, it seems
relatively probable that at any given time some stellar BHs in the
Galactic centre are bright X-ray sources as they accrete from the
clumpy interstellar gas (poster by Patrick Deegan in this conference;
Nayakshin \& Sunyaev, in preparation).

As for EMRI rates and properties, the determination of how 2-body
relaxation shapes the stellar distribution around the MBH is only a
first --crucial-- step. A robust estimate of the fraction of stars
that eventually inspiral into LISA band, rather than plunge directly
through the horizon while still on a wide orbit
\cite{HB95,HA05} will probably require the development of a specific
code. For stars on very eccentric orbits, one needs to follow the
combined effects of GW emission and relaxation on a timescale
significantly shorter than allowed by the present MC code. Recently
Hopman \& Alexander (\cite{HA06} and these proceedings) have
considered, for the first time in the study of EMRIs, the role of
``resonant relaxation'', i.e., of the random changes in eccentricity
and orientation of the orbital planes due to the non-vanishing but
fluctuating torque exerted on an orbit by the other orbits, each
considered as an elliptical mass wire
\cite{RT96}. These authors find that resonant relaxation can increase 
the EMRI rate by of order a factor 10, an exciting result which is
calling for confirmation by other computation techniques. Resonant
relaxation can be included in the MC algorithm using the same
approximate formulae as Hopman \& Alexander. Unfortunately, this
treatment relies on the calibration realised through a few very
low-$N$ $N-$body simulations \cite{RT96} and more accurate numerical studies of
this potentially important effect are called for (see Touma, these
proceedings for a possible high-efficiency method).

\ack

PAS is indebted to Holger Baumgardt for invaluable help with the
modification and use of {\NBFOUR}. We thank him for discussions and
making available to us the data from his $N-$body simulations. We are
grateful to Tal Alexander, John Fregeau, Atakan G\"urkan and Frederic
Rasio for enlightening discussions. MF received invaluable
explanations about statistical tests for the distribution of X-ray
sources around {\SgrA} from Laurent Eyer.

The $N-$body simulations of 2-component models were performed on the
{\em GRAPE\/} computers at the Astronomisches Rechen Institut in
Heidelberg, thanks to the support of Rainer Spurzem, Peter Berczik and
Peter Schwekendiek.

This research was supported by NASA ATP Grant NAG~5-13229. MF and VK
acknowledge the Aspen Center for Physics where work on this project
has been partially carried out, during the workshop ``LISA Data:
Analysis, Source and Science''. The participation of MF in
this workshop was supported in part by NASA grant NNG05G106G. The work
of MF in Cambridge is funded through the PPARC rolling grant to the
theory group at the IoA. The work of PAS has been supported in the
framework of the Third Level Agreement between the DFG (Deutsche
Forschungsgemeinschaft) and the IAC (Instituto de Astrof\'\i sica de
Canarias).


\end{document}